\documentclass[%
 reprint,
 aip,
showkeys,
 amsmath,amssymb,
longbibliography,
]{revtex4-1}

\usepackage{graphicx}
\usepackage[caption=false]{subfig} 
\usepackage{pifont}
\usepackage{accents} 
     
\usepackage{dcolumn}
\usepackage{bm}

\newcommand{\ee}{\end{equation}} 
\newcommand{\be}{\begin{equation}}

\usepackage[mathscr]{euscript}  
\usepackage{xcolor}  

\usepackage{tcolorbox}
\usepackage{comment}
\usepackage{float}

\makeatletter
\newsavebox{\@brx}
\newcommand{\llangle}[1][]{\savebox{\@brx}{\(\m@th{#1\langle}\)}%
  \mathopen{\copy\@brx\kern-0.5\wd\@brx\usebox{\@brx}}}
\newcommand{\rrangle}[1][]{\savebox{\@brx}{\(\m@th{#1\rangle}\)}%
  \mathclose{\copy\@brx\kern-0.5\wd\@brx\usebox{\@brx}}}
\makeatother

\makeatletter
\newcommand{\vast}{\bBigg@{3.4}}
\makeatother

\usepackage{tikz}

\makeatletter
\newcommand\emailx[1]{%
\move@AF%
\def\@affil{{\normalfont\,#1\strut}{}}%
}%

\begin{document}

\preprint{ApS/123-QED}

\title{Quadrupolar gyration of a Brownian particle in a confining ring}

\author{Iman Abdoli}
\email{iman.abdoli@hhu.de}
\affiliation{Institut für Theoretische Physik II - Weiche Materie, Heinrich-Heine-Universität Düsseldorf, Universitätsstraße 1, D-40225 Düsseldorf, Germany}


\author{Hartmut L\"{o}wen}
\affiliation{Institut für Theoretische Physik II - Weiche Materie, Heinrich-Heine-Universität Düsseldorf, Universitätsstraße 1, D-40225 Düsseldorf, Germany}

\begin{abstract} 
We develop a minimal theoretical model that reveals a structured steady-state flux field with  four alternating local circulation, a phenomenon we refer to as \textit{quadrupolar gyration}. 
A passive Brownian particle is confined to move in a ring-shaped trap and driven far from equilibrium solely by anisotropic thermal fluctuations from two orthogonal heat baths held at different temperatures. By breaking detailed balance, this fundamental temperature anisotropy induces a robust nonequilibrium steady state characterized by probability currents of the particle's motion. Remarkably, these currents self-organize into a distinctive quadrupolar vortex pattern, providing a clear signature of emergent symmetry breaking, irreversible entropy production, and coherent motion in minimal passive systems.	
	

\end{abstract}


\maketitle

Brownian gyrator is a minimal microscopic system that elegantly demonstrates how thermal anisotropy--where the particle’s orthogonal spatial degrees of freedom are  coupled to heat baths at unequal temperatures-- can inherently generate steady-state rotational fluxes without external forces or torques~\cite{filliger2007brownian, ciliberto2013heat,  dos2021stationary, movilla2021energy, squarcini2022fractional, cerasoli2022spectral, siches2022inertialess}. Originally conceived as a nanoscale heat engine consisting of a particle diffusing in a two-dimensional anisotropic harmonic potential with misaligned principal axes relative to temperature axes, the Brownian gyrator transforms thermal noise from orthogonal heat baths into persistent probability currents, exemplifying symmetry breaking and directed motion in microscopic systems.~\cite{miangolarra2022thermodynamic, abdoli2022escape, miangolarra2024energy, miangolarra2024minimal, miangolarra2024stochastic}.
This mechanism has since emerged as a fundamental model system for probing its electrical counterpart, universal trade-offs governing nonequilibrium engines~\cite{chiang2017electrical, Pietzonka:2018utp} and for studying inertia effect, trajectory-level fluctuations, and entropy production through stochastic thermodynamic tools~\cite{bae2021inertial, seifert2012stochastic, dotsenko2013two, koyuk2020thermodynamic}. 
The gyrator has also been reinterpreted as a minimal information engine, where work extraction is viewed as internal information flow, akin to a Maxwell demon~\cite{leighton2024information}.


The emergence of gyrating motion crucially depends on spatial cross-correlations, which can be originated internally from potential structures or externally from magnetic fields~\cite{abdoli2020correlations, abdoli2022tunable, muhsin2025active, adersh2025active}. Recent work reveals nonharmonic double-well potentials induce counter-rotating fluxes near minima, highlighting geometry's control over nonequilibrium current patterns~\cite{chang2021autonomous}. The theoretical predictions of a Brownian gyrator have been realized in experiments by emulating anisotropic thermal noise via strongly fluctuating electric fields applied along one direction, effectively mimicking the role of additional temperature~\cite{argun2017experimental}.
The efficiency of an engine relies sensitively on how external mechanical loads, such as shear forces, are applied and tuned to extract work from the engine~\cite{abdoli2025enhanced}. 


Unlike the conventional two-dimensional anisotropic harmonic potential—where misaligned principal axes drive dynamics—a ring-shaped confinement imposes radical geometric restrictions, pinning particles to a fixed radius while enabling free angular diffusion. This transforms the system’s essential character from planar stochastic motion to quasi-one-dimensional gyration governed by curvature and confinement. In this geometry, the particle's motion becomes predominantly angular, and any remaining radial diffusion is tightly suppressed. This shift   introduces qualitatively different modes of flux formation. Consequently, the ring trap provides a minimalistic and highly controllable platform for investigating nonequilibrium steady states driven by anisotropic fluctuations, and highlights opportunities to explore the roles of spatial geometry and thermal noise in generating rotational probability currents.

Optical ring traps have enabled the controlled realization of persistent circulating currents, noise rectification, and symmetry breaking under anisotropic driving~\cite{roichman2008influence, sun2009brownian, saha2024cybloids}. Recent experiments on optically confined colloids have demonstrated that temperature gradients can drive sustained particle transport via optothermal hydrodynamic coupling, elucidating how ring geometries mediate collective phenomena such as unidirectional circulation and spontaneous vortex formation~\cite{chand2025optothermal}. Theoretical studies have examined Brownian motion on a ring in diverse contexts, including winding statistics~\cite{kundu2014winding}, stochastic resetting~\cite{grange2022winding}, and large deviation theory~\cite{proesmans2019large}. In active matter, confined self-propelled particles exhibit curvature-sensitive accumulation and current reversal~\cite{fazli2021active}, highlighting the geometric control of transport in driven systems.

Here we study a paradigmatic stochastic system: a Brownian particle confined to a narrow ring and coupled to two orthogonal heat baths at unequal temperatures. The resulting interplay between thermal anisotropy and geometric curvature drives the emergence of steady-state probability currents and entropy production, even in the absence of external forces or torques. Remarkably, we find that the system spontaneously generates a steady-state flux field with four alternating vortical sectors, forming a robust pattern that we term \textit{quadrupolar gyration}.
These nonequilibrium features are captured analytically via a perturbative solution of the Fokker–Planck equation in the narrow-ring limit, and validated by direct numerical simulations. We compute the full spatial structure of the steady-state probability distribution, probability currents, and entropy production density, and reveal how their topology and intensity depend sensitively on the noise anisotropy and ring radius.

\section*{Results}

\begin{figure}
	\centering
	\includegraphics[width=8cm]{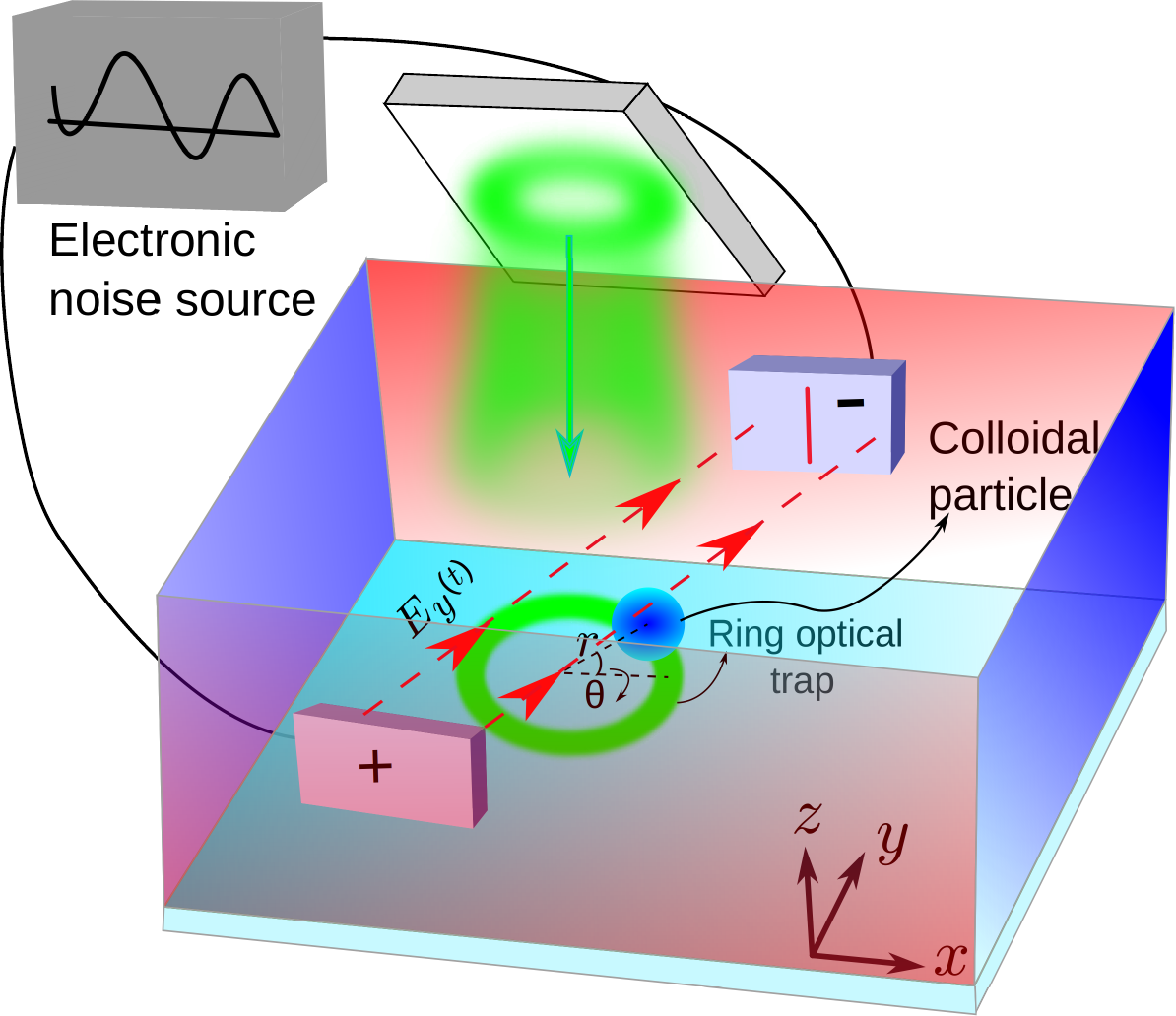}
	\caption{ \textbf{Schematic illustration of the model and the experimental realization.}  
		A colloidal particle (blue) is confined within a ring-shaped optical trap generated via a spatial light modulator. Anisotropic thermal noise is introduced by applying a fluctuating electric field $E_y(t) \sim$ white noise along the $y$-direction using a pair of electrodes placed above and below the trap. This creates an effective hot bath along $y$, while the ambient aqueous medium provides a colder bath along $x$. Note that the system is two-dimensional, and the three-dimensional schematic is for clarity of illustration.}
	\label{figure01}
\end{figure}

\textbf{Model and setup}
We consider a Brownian particle restricted to move in a narrow ring-shaped confinement, subjected to anisotropic thermal fluctuations. The particle resides in a two-dimensional plane and experiences independent noise sources with distinct temperatures \(T_x\) and \(T_y\) along the orthogonal Cartesian axes $x$ and $y$ (see Fig.~\ref{figure01}). Unlike conventional Brownian gyrators confined in planar harmonic wells, here the particle is radially confined to a circular shell of radius \(R\), effectively reducing the accessible degrees of freedom to quasi-one-dimensional motion along the angular direction. This geometry eliminates the possibility of global translation or drift, isolating rotational degrees of freedom and enabling clear observation of emergent circulating currents.

The confinement is modeled by a radial potential of the form
\begin{equation}
	U(r) = \frac{k}{2}(r - R)^2,
\end{equation}
which restricts fluctuations around the mean radius \(R\) with stiffness \(k\) and distance $r=\sqrt{x^2+y^2}$, while allowing free diffusion in the azimuthal direction. 

To analyze the system, we derive the corresponding Fokker–Planck equation in polar coordinates and study the emergence of steady-state solutions under the narrow-ring approximation, \(\sqrt{T_x/k} \ll R\), where radial confinement is tight but finite. The probability density \(P(r,\theta,t)\) evolves according to the Fokker--Planck equation, expressed in compact tensorial form as (see Supplementary Material (SM))

\begin{equation}
	\label{eq:FPE_tensor}
	\frac{\partial P}{\partial t} = - \nabla \cdot \left[ \mu \mathbf{F} P - \mathbf{D}(\theta) \cdot \nabla P \right],
\end{equation}
where \(\mathbf{F} = -\nabla U(r) = -k(r - R)\, \hat{\mathbf{e}}_r\) is the radial restoring force derived from the ring-shaped potential. As we derive in the SM, the diffusion coefficient picks up a tensorial form in polar basis which reads


\begin{equation}
	\label{eq:Dtensor}
	\mathbf{D}(\theta) = \mu T 
	\begin{pmatrix}
		1 - \alpha \cos 2\theta & \alpha \sin 2\theta \\
		\alpha \sin 2\theta & 1 + \alpha \cos 2\theta
	\end{pmatrix}.
\end{equation}

Here, \(\mu=\gamma^{-1}\) is the mobility with $\gamma$ being the friction coefficient, \(\alpha = (\beta - 1)/(\beta+ 1)\) is the dimensionless anisotropy parameter and \(T = T_x(\beta + 1)/2\) is the mean effective temperature with \(\beta = T_y/T_x\) where \(\beta=1\) corresponds to the isotropic system. We set the Boltzmann constant $k_B$ to unity throughout this work. This tensorial formulation naturally leads to the full Fokker--Planck equation in polar coordinates as presented in the Methods and SM. \\




\begin{figure*}[t]
	\centering
	\includegraphics[width=16.5cm]{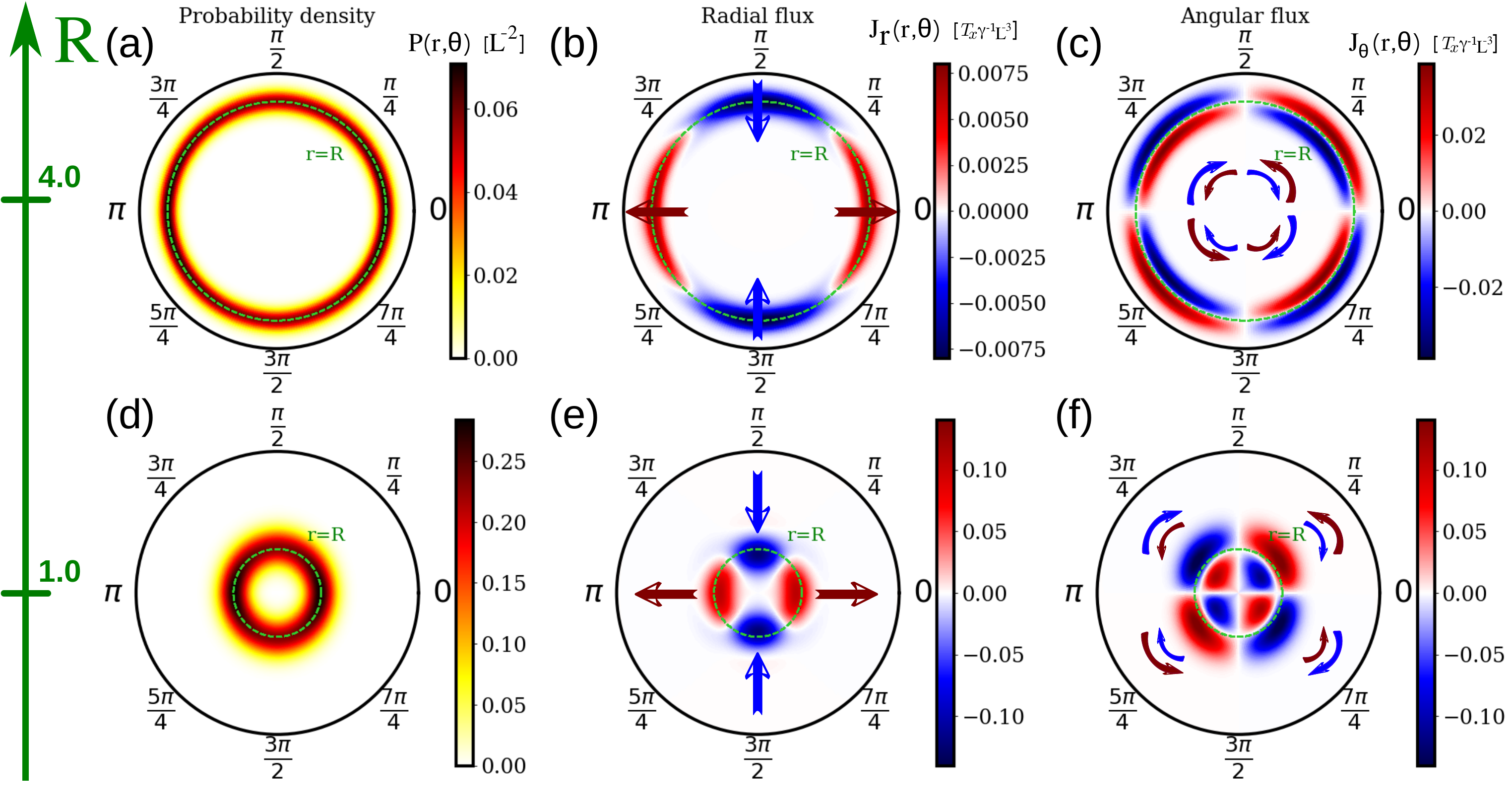}
	\caption{
		\textbf{Steady-state probability density and probability flux components.}
		Panels (a–c) correspond to \(R = 4.0\), while (d–f) show \(R = 1.0\). In all cases, the trap stiffness is \(k = 20\), and the temperature ratio \(\beta=1.5\), with characteristic length scale \(L = \sqrt{T_x / k}\).	
		(a,d) The probability distribution, plotted from Eq.~\eqref{eq:PDF}, is sharply localized around the ring \(r = R\) (indicated by dashed circles), and anisotropic thermal noise induces an angular modulation visible as slight distortions along the \(\theta\)-direction.	
		(b,e) The radial flux, plotted from Eq.~\eqref{radial_flux_methods}, highlights the local circulating currents that maintain the non-equilibrium steady state. Inward and outward flow regions are indicated by blue and dark-red straight arrows, respectively, illustrating the direction of radial motion.	
		(c,f) The angular flux from Eq.~\eqref{angular_flux_methods}, emphasizes the sustained azimuthal circulating motion characteristic of gyrators. Blue and dark-red curved arrows indicate clockwise  and counter-clockwise  circulations, respectively.	
		For \(R = 4.0\) (top row), the probability density forms a broad annulus with smoother flux patterns, indicating a more delocalized state. In contrast, for \(R = 1.0\) (bottom row), the probability is tightly localized and the flux components are more concentrated, highlighting stronger confinement and enhanced rotational features.	
		The dashed circles at \(r = R\) serve as a reference to the nominal ring radius and emphasize the degree of spatial localization.
	}
	
	\label{figure02}
\end{figure*}

\textbf{Experimental realization} 
An experimental realization of a ring-constrained Brownian gyrator can be achieved by combining a \textit{ring-shaped optical trap}, as implemented by Chand \textit{et al.}~\cite{chand2025optothermal}, with the \textit{anisotropic thermal noise generation technique} developed by Volpe \textit{et al.}~\cite{argun2017experimental}. The system consists of a colloidal particle confined to a narrow optical ring trap, generated via a spatial light modulator (SLM) that sculpts the laser intensity profile into a circular minimum in the horizontal \( x\text{--}y \) plane. This provides strong radial confinement while allowing free azimuthal motion, effectively localizing the particle to a ring of radius \( R \) immersed in water (see Fig.~\ref{figure01}).

To break thermal equilibrium and introduce direction-dependent noise, a stochastic electric field \( E_y(t) \) is applied along the vertical \( y \)-axis using a pair of microelectrodes embedded above and below the sample. The electrodes are driven by a white-noise voltage source, creating a high-frequency, band-limited fluctuating field that couples to the charged particle. This mimics an effective hot thermal bath along \( y \) by enhancing stochastic forcing in that direction. In contrast, the ambient aqueous environment maintains a lower effective temperature along \( x \). The resulting anisotropy \( T_y > T_x \), combined with circular confinement, drives the system into a genuine nonequilibrium steady state exhibiting persistent circulating probability currents.\\

\textbf{Probability density and flux structure}
To characterize the non-equilibrium steady state established by anisotropic thermal driving in a ring trap, we analyze the stationary probability distribution and associated probability currents. Unlike planar Brownian gyrators, where fluxes can form simple rotational loops, the ring geometry imposes a strong radial confinement that fundamentally reshapes both the spatial distribution and the flux topology. In this quasi-one-dimensional setting, the angular direction remains free, but the anisotropic diffusion induces a rich angular modulation and persistent circulating currents. The resulting structure reflects a delicate interplay between confinement strength, thermal anisotropy, and spatial geometry.

The steady-state probability density \( P(r, \theta) \) is obtained from the corresponding Fokker--Planck equation, given in Eq.~\eqref{eq:FPE_tensor} and Eq.~\eqref{eq:Dtensor}, under the narrow-ring approximation, where the radial width is small but finite. This allows for a perturbative treatment as presented in the Supplementary Material. At leading order, the radial distribution is Gaussian around the trap radius \( r = R \), with an angular modulation governed by the anisotropy parameter \( \alpha \). The approximate form of the steady-state solution reads
\begin{equation}
	\label{eq:PDF}
	P(r, \theta)\! \approx\! \frac{1}{\mathcal{N} \sqrt{T_r(\theta)}} \!\exp\!\left[\!-\frac{k \rho^2}{2 T_r(\theta)}\! -\! \alpha\! \left(a_1 \rho \!+\! a_2 \rho^2\right)\! \cos(2\theta)\! \right],
\end{equation}
where the effective radial temperature is
\(
T_r(\theta) = T (1-\alpha\cos(2\theta))
\), \( \rho = r - R \), and \( \mathcal{N} \approx 2\pi R \sqrt{2\pi/k} \) is the normalization constant. The geometric coefficients appearing in the angular modulation are given by
\(a_1 = (2kR^2 - 4T)/(kR^3 + RT)\) and \( a_2 = (-2kR^2 + T)/(kR^4 + R^2T)\). The key feature of this expression is the angular modulation of the density, which breaks rotational symmetry and encodes the directionality of fluxes. The radial and angular fluxes can be calculated using the probability density in Eq.~\eqref{eq:PDF} as shown in Methods and SM. 

\begin{figure}[t]
	\centering
	\includegraphics[width=8.0cm]{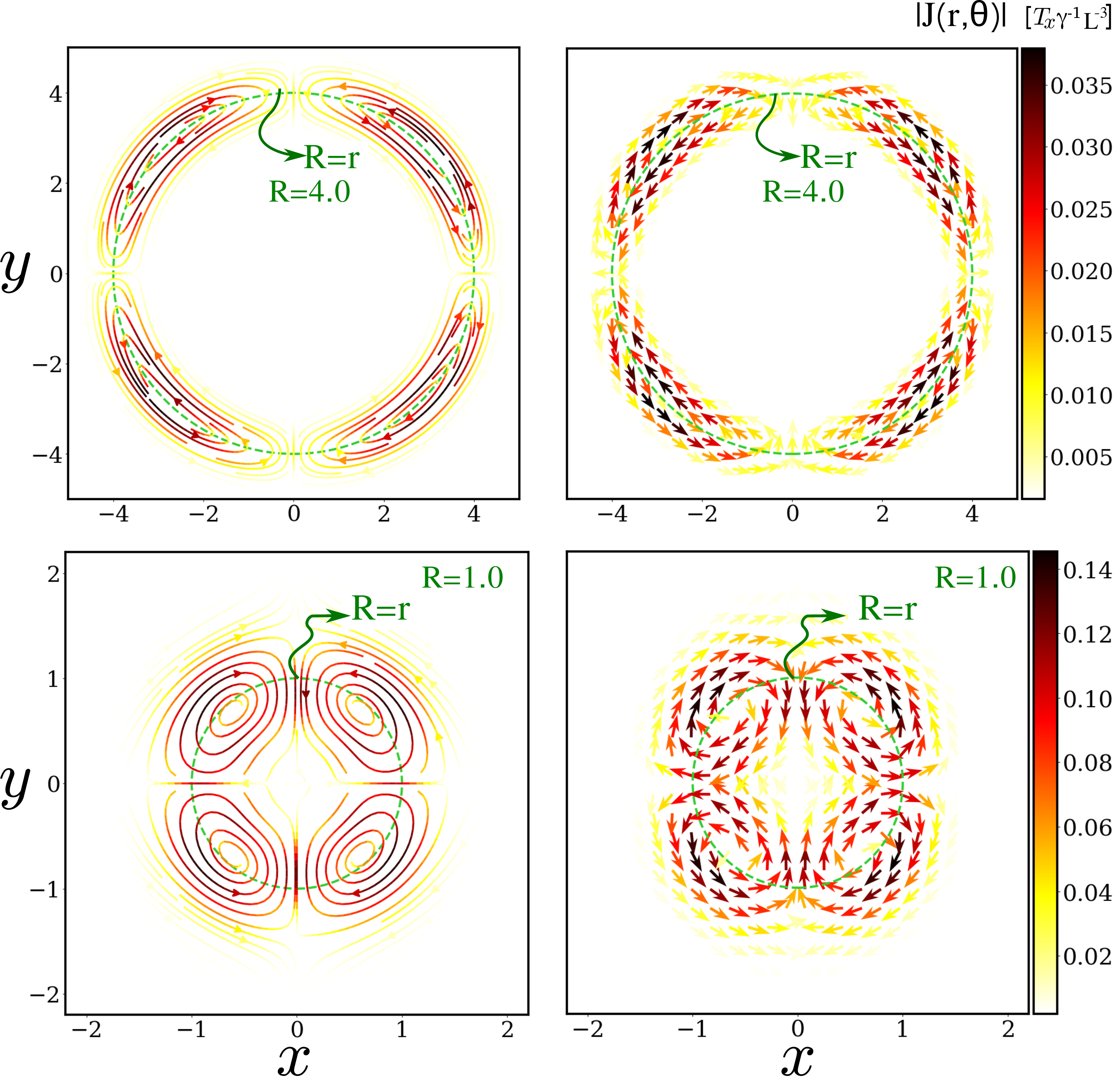}
	\caption{
		\textbf{Probability flux streamlines and quiver plots.}
		Top panels correspond to $R = 4.0$, bottom panels to $R = 1.0$. The steady-state probability flux $\mathbf{J}(r,\theta)$ is projected in Cartesian coordinates $(x, y)$, with streamlines visualizing circulation patterns and quiver arrows indicating local flux directions. Color maps encode the flux magnitude $|\mathbf{J}(r,\theta)|$ for \(\beta=1.5\) and \(k=20\), with $L = \sqrt{T_x/k}$. For $R = 4.0$ (top), the flux forms extended vortex-like loops, while for $R = 1.0$ (bottom), circulation becomes more localized with pronounced radial and azimuthal components due to tighter confinement and stronger anisotropy. Green dashed circles mark the nominal ring position $r = R$. Results computed directly from Eqs.~\eqref{radial_flux_methods}--\eqref{angular_flux_methods}. 
	}
	
	\label{figure03}
\end{figure}

As shown in Fig.~\ref{figure02}, the steady-state density forms a sharply localized ring around \( r = R \), consistent with the imposed confinement. The modulation along \( \theta \) is subtle but discernible, reflecting the influence of anisotropic diffusion. The associated flux field, shown in both polar and Cartesian representations (Figs.~\ref{figure02}--\ref{figure03}), exhibits a striking quadrupolar pattern. Radial fluxes, inward and outward (Fig.~\ref{figure02}b,e), coexist with angular circulations (Fig.~\ref{figure02}c,f), generating a coherent vortex structure. The full streamline plots (Fig.~\ref{figure03}) highlight this behavior, with extended vortical loops for larger \( R \), and more compact, tightly wound recirculations for smaller \( R \). This transition indicates enhanced localization and stronger angular anisotropy at small radii, illustrating how trap geometry governs the qualitative features of the steady state.

\textbf{Entropy production density.}
We now quantify local irreversibility in the non-equilibrium steady state, by evaluating the spatially resolved entropy production density~\cite{puglisi2025brief, fernandez2024nonequilibrium, cates2022stochastic, loos2019heat} $\sigma(r,\theta)$, defined as $\sigma(r,\theta) = \mathbf{J}^\top \mathbf{D}^{-1} \mathbf{J} / P(r,\theta)$, where $\mathbf{J}$ is the steady-state flux, $\mathbf{D}$ is the diffusion tensor in Eq.~\eqref{eq:Dtensor}, and $P(r,\theta)$ is the steady-state probability density as in Eq.~\eqref{eq:PDF}. In our system, this quantity admits an explicit analytic expression under the narrow-ring approximation due to the constant determinant of $\mathbf{D}$ (see SM). The result exhibits a pronounced quadrupolar angular modulation through $\cos^2(2\theta)$ and $\sin^2(2\theta)$ terms, modulated by a radial Gaussian envelope centered around the trap minimum $r = R$.

The radial entropy production profile \( \sigma(r) \), obtained via angular averaging, reveals a non-monotonic structure that depends sensitively on the ring radius \( R \) (Fig.~4). For larger rings, entropy production spreads broadly along the radial direction, while smaller rings concentrate dissipation more sharply near the center. In both cases, \( \sigma(r) \) develops symmetric off-centered peaks and can exhibit a suppression at \( r = R \), despite the maximum of the probability density there. This counterintuitive dip reflects the vanishing of radial flux and the geometric modulation of angular currents near the trap minimum. The overall amplitude of \( \sigma(r) \) scales as \( \alpha^2 \), underscoring that irreversibility emerges purely from the imposed anisotropic fluctuations.

\begin{figure}[t]
	\centering
	\includegraphics[width=8.3cm]{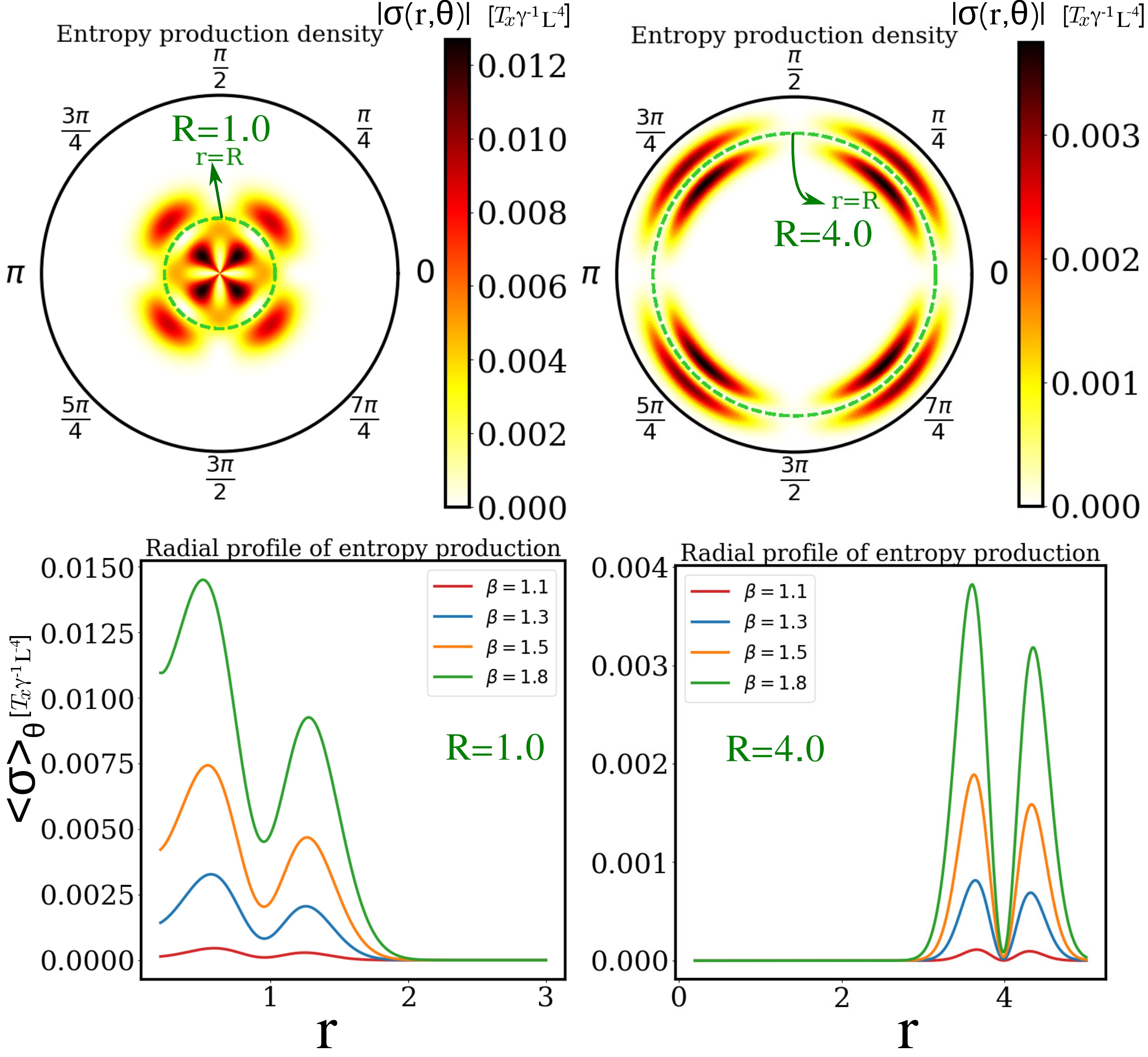}
	\caption{
\textbf{Entropy production density and radial profiles.}  
Top: Steady-state entropy production density \(|\sigma(r,\theta)|\) plotted in polar coordinates for two ring radii, \(R = 1.0\) (left) and \(R = 4.0\) (right). System parameters are \(k = 20\), \(\beta = 1.5\), and characteristic length scale \(L = \sqrt{T_x/k}\). The dashed green circle denotes the nominal trap location \(r = R\). For the smaller radius, dissipation is strongly concentrated near the center, forming localized hot spots due to enhanced angular modulation and curvature effects. In contrast, for the broader ring, entropy production becomes more azimuthally extended and radially delocalized, forming a narrow quadrupolar band.  
Bottom: Radial profiles of the angularly averaged entropy production \(\langle \sigma(r,\theta) \rangle_\theta\) for varying anisotropy \(\beta = T_y/T_x\), shown for both trap sizes. Increasing \(\beta\) amplifies the overall dissipation and shifts it outward, consistent with enhanced rotational driving. All curves exhibit non-monotonic structure with symmetric off-centered peaks around the trap minimum, reflecting the geometric suppression of flux at \(r = R\).	Results obtained from the analytical expression Eq.~\eqref{EPD_methods} and Eq.~\eqref{radial_EPD_methods}.}
	
	\label{figure04}
\end{figure}

\textbf{Quadrupolar gyration}
The emergence of quadrupolar circulation in the ring Brownian gyrator arises from the interplay between confinement geometry and anisotropic thermal noise. Although the particle is confined to a rotationally symmetric ring, the projection of the diffusion anisotropy onto the local polar basis introduces an effective angular modulation of noise. Specifically, the effective temperatures in the radial and angular directions vary with position as \(
T_r(\theta) = T (1-\alpha\cos(2\theta))
\) and \(
T_\theta(\theta) = T (1+\alpha\cos(2\theta))
\). This spatial variation in local diffusivity generates alternating thermodynamic forces along the ring, leading to localized circulating currents in the steady state.

The underlying diffusion tensor, given in Eq.~\eqref{eq:Dtensor}, has eigenvalues \( \lambda_- = T_x \) and \( \lambda_+ = T_y \), corresponding to the directions of minimal (cold) and maximal (hot) axis, respectively. While the eigenvalues are uniform, the associated eigenvectors rotate continuously along the ring: the hot axis points along \( \mathbf{v}_{\text{hot}}(\theta) = (\sin\theta, \cos\theta) \), while the cold axis lies along \( \mathbf{v}_{\text{cold}}(\theta) = (\cos\theta, -\sin\theta) \). In polar coordinates, the hot axis is locally aligned with angle \( \phi = \theta \), and the cold axis with \( \phi = -\theta \), indicating a continuously rotating anisotropy tied to the particle’s position.

\begin{figure}
	\centering
	\includegraphics[width=8.2cm]{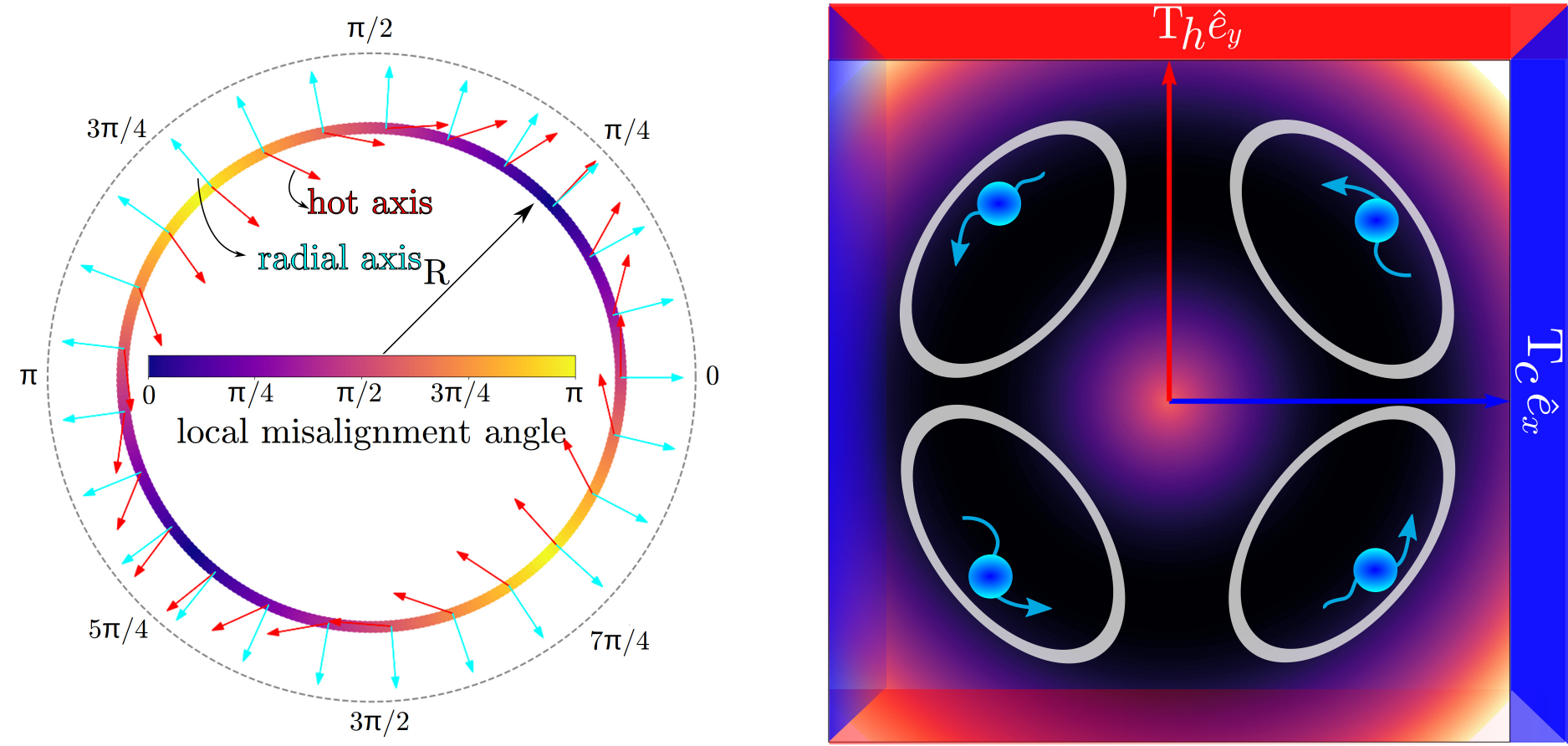}
	\caption{ \textbf{Quadrupolar flux structure from locally rotating hot axes.} 
	(Left) Visualization of the local misalignment angle (color) between the radial direction and the hot axis of anisotropic diffusion as a function of position along the ring. Blue and red arrows represent the cold and hot diffusion directions, respectively, highlighting the continuous spatial rotation of the anisotropic diffusion tensor.
	(Right) Schematic representation of the resulting quadrupolar current pattern: the ring can be heuristically decomposed into four quadrants, each behaving as an effective Brownian gyrator with locally projected anisotropy. The angular variation of the hot axis generates alternating clockwise (CW) and counterclockwise (CCW) circulating fluxes, collectively forming a robust quadrupolar current structure in the steady state. 
	} 
	\label{figure05}
\end{figure}

This spatially rotating anisotropy breaks detailed balance and induces a structured non-equilibrium current pattern, as illustrated in Fig.~~\ref{figure05}. The ring can be heuristically viewed as partitioned into four quadrants, each effectively acting as a local Brownian gyrator. In each region, the radial confinement projects the anisotropic fluctuations into misaligned geometric directions, generating alternating clockwise and counterclockwise circulating fluxes that collectively form the global quadrupolar pattern.

\section*{Conclusion}
Our study reveals how spatial confinement and thermal anisotropy can conspire to generate intricate nonequilibrium dynamics in an otherwise force-free, passive system. By coupling a Brownian particle to orthogonal heat baths within a ring-shaped trap, we uncover a robust mechanism for producing quadrupolar probability currents and structured entropy production—without invoking self-propulsion, external fields, or interactions. The ring geometry plays a central role, transforming what would otherwise be linear or planar transport into circulating steady-state motion characterized by symmetry breaking and vorticity inversion.

Unlike planar Brownian gyrators, where rotation emerges from the misalignment of the  potential axes relative to anisotropic noise, the ring confinement forces the system into a quasi-one-dimensional topology where angular and radial fluctuations become tightly coupled. This gives rise to steady-state flux field with four alternating gyrating sectors, i.e. quadrupolar gyration, and entropy production patterns that are highly localized yet globally balanced. Our analytical treatment captures these features in the narrow-ring limit, while simulations confirm their persistence across a broader range of parameters.

Beyond its theoretical significance, our model offers a versatile framework for designing stochastic thermal machines~\cite{whitelam2023demon} where geometry and temperature anisotropy govern microscopic motion~\cite{rings2012rotational}. 
It opens avenues for exploring fluctuation-driven organization~\cite{manikandan2019efficiency} in confined colloidal systems~\cite{williams2013direct, antonov2025controlling, cereceda2023overcrowding, lutz2004single, wei2000single} including active matter~\cite{fodor2022irreversibility, dolai2022inducing}, with implications for microrheology, multi-terminal stochastic heat engines~\cite{netz2020approach}, transport in curved environments~\cite{villada2021single}, and synthetic engines. Our findings provide a basis for exploring coupled ring gyrators to explore collective flux organization and energy transfer, which could lead to synchronization, phase-locking, or emergent collective currents. Future directions include designing optimal loading protocols for work extraction and exploring whether similar quadrupolar flux patterns can arise in chiral (passive or active) particles with spatially varying handedness.


\section*{Methods}
\label{Methods}

\textbf{Theoretical predictions}
We analytically characterize the non-equilibrium steady state of a Brownian particle confined in a narrow ring and subject to anisotropic thermal fluctuations. The system is described by the time-dependent Fokker--Planck equation in polar coordinates \((r, \theta)\), which governs the evolution of the probability density \(P(r, \theta, t)\). The dynamics involve coordinate-dependent drift and diffusion due to both confinement and the angular dependence of the diffusion tensor induced by temperature anisotropy.

As derived in the SM, the time-dependent Fokker--Planck equation takes the divergence form 
\begin{equation}
	\partial_t P(r, \theta, t) = -\frac{1}{r} \frac{\partial}{\partial r} \left[ r J_r(r, \theta, t) \right] - \frac{1}{r} \frac{\partial}{\partial \theta} J_\theta(r, \theta, t),
\end{equation}

where the flux components \(J_r\) and \(J_\theta\) follow from direct projection of the anisotropic Cartesian flux onto the polar basis, which read (see SM for full derivation)
\begin{align}
	J_r &= -\mu k(r - R) P - D_r(\theta) \frac{\partial P}{\partial r} - \frac{D_{r\theta}(\theta)}{r} \frac{\partial P}{\partial \theta}, \\
	J_\theta &= - D_{r\theta}(\theta) \frac{\partial P}{\partial r} - \frac{D_\theta(\theta)}{r} \frac{\partial P}{\partial \theta},
\end{align}
with the angle-dependent diffusion coefficients defined as
\begin{align}
	D_r(\theta) &= \mu T(1 - \alpha \cos(2\theta)), \\
	D_\theta(\theta) &= \mu T(1 + \alpha \cos(2\theta)), \\
	D_{r\theta}(\theta) &= \mu T \alpha \sin(2\theta),
\end{align}
where  \(\mu\) is the mobility, \(\alpha = (\beta - 1)/(\beta+ 1)\) is the dimensionless anisotropy parameter and \(T = T_x(\beta + 1)/2\) is the mean effective temperature with \(\beta = T_y/T_x\) where \(\beta=1\) corresponds to the isotropic system. 


To make analytical progress, we perform a perturbative expansion valid in the narrow-ring limit around $r = R$. Retaining terms only linear in the anisotropy parameter $\alpha$, we obtain approximate expressions for the steady-state probability flux components:
\begin{align}
	\label{radial_flux_methods}
	J_r(r,\theta) 
	&\approx \frac{\alpha\mu \sqrt{kT}\,e^{-\frac{k(r-R)^2}{2T}}}{(2 \pi)^{3/2}R}
	\left(a_1 + 2 a_2 (r - R)\right) \,\cos(2\theta), 
	\\
	J_\theta(r,\theta) 
	&\approx \frac{\alpha\mu\sqrt{k}\,e^{-\frac{k(r-R)^2}{2\mu T}}}{\sqrt{T}(2 \pi)^{3/2}R\,r}
	\Bigl(T + \notag \\
	\quad & (r - R)\bigl[k R - 2 (a_1 + a_2 (r - R))T \bigr]\Bigr)
	\sin(2\theta).
	\label{angular_flux_methods}
\end{align}
where the coefficients $a_1$ and $a_2$ depend only on $k$, $R$, and $T$, and are given by:
\begin{align}
	a_1 &= \frac{2k R^2 - 4 T }{k R^3 + R T },  \\
	a_2 &= \frac{-2k R^2 + T }{k R^4 + R^2 T }. 
\end{align}

These expressions capture the emergence of quadrupolar structure in the steady-state fluxes, driven solely by anisotropic fluctuations in a rotationally symmetric trap. In this limit, the probability density remains approximately Gaussian in the radial direction, while the angular modulation of the fluxes generates circulating currents that break detailed balance. This perturbative framework provides a controlled route to predicting flux patterns, entropy production, and irreversibility in confined nonequilibrium systems.

At leading order in the narrow-ring limit, the steady-state probability and flux fields admit approximate analytical expressions, yielding a closed-form for the local entropy production density
\begin{align}
	\label{EPD_methods}
	&	\sigma(r,\theta) 
	= \notag \\ 
	& \frac{\alpha^2 \, e^{- \frac{k (r - R)^2}{2T}} \, \sqrt{k} \, \mu }
	{4 \sqrt{2}\,\pi^3 \, T^{3/2} \, (R - R \alpha^2)}
	\Bigl[
	\bigl(a_1 + 2 a_2 (r - R)\bigr)^2 T^2 \cos^2(2\theta)
	\nonumber\\
	&
	+ \frac{\bigl(T \!+\! (r\! -\! R)\bigl(kR\! -\! 2(a_1\! +\! a_2 (r\! -\! R))T\bigr)\bigr)^2 \sin^2(2\theta)}
	{r^2}
	\Bigr],
\end{align}
which exhibits a pronounced quadrupolar angular structure encoded by $\cos^2(2\theta)$ and $\sin^2(2\theta)$ modulations, and a radial Gaussian envelope centered around $r = R$.

The corresponding radial profile, obtained by angular averaging, reads
\begin{align}
	\label{radial_EPD_methods}
	\sigma(r) =\ 
	& \frac{\alpha^2 e^{-\frac{k (r - R)^2}{2 T}} \sqrt{k} \mu}
	{8 \sqrt{2} \pi^4 T^{3/2} (R - R \alpha^2)} \left[
	\pi (a_1 + 2 a_2 (r - R))^2 T^2 \right. \notag \\
	& \left. +\ \frac{\pi \left( T\! +\! (r\! -\! R)(k R\! -\! 2(a_1\! +\! a_2 (r\! -\! R)) T) \right)^2}{r^2}
	\right],
\end{align}
revealing nontrivial radial dependence with peaks that shift and grow with increasing anisotropy $\alpha$. The amplitude of $\sigma(r)$ scales as $\alpha^2$, highlighting that entropy production is entirely driven by the imposed temperature gradient. This structure reflects the interplay between geometric confinement, flux topology, and anisotropic fluctuations.\\

\textbf{Numerical simulations}
To verify our analytical predictions, we also perform Brownian dynamics simulations of the system. The dynamics of the particle's position \(\bm{r}(t) = (x(t), y(t))\) can be described by the following overdamped Langevin equation

\begin{align}
	\dot{\bm{r}} = -\mu \bm{\nabla} U(r) + \bm{\xi}(t)
\end{align}
where \(\mu\) is the mobility, \(\boldsymbol{\xi}(t) = (\xi_x(t), \xi_y(t))\) is a Gaussian white noise vector with zero mean and \(\langle \eta_i(t)\eta_j(t') \rangle = 2\mu T_i \delta_{ij} \delta(t - t')\) where \(i,j \in \{x, y\}\)

The simulations are performed using an Euler-Maruyama integration scheme with a discrete time step \(d t=1\times 10^{-3}\gamma/k\), chosen to be sufficiently small to ensure numerical stability and accuracy. Thermal noise is implemented by drawing random numbers from a normal distribution at each time step. 

While no simulation results are shown in the main text, all theoretical predictions have been cross-validated against Brownian dynamics simulations in the Supplementary Material. In the narrow-ring regime where the analytical approach is controlled, we find an excellent agreement.

\section*{DATA AVAILABILITY}
The data that support the findings of this study are available
from the corresponding author upon reasonable request.

\section*{Code AVAILABILITY}
The source code of the Brownian dynamics simulations is available from the corresponding author on reasonable request.


\section*{References}

\begin{acknowledgements}
This work was funded by the Deutsche Forschungsgemeinschaft (DFG, German Research Foundation) under project number 556762905 — AB 1083/1-1.

	
\end{acknowledgements}

\subsection*{Author contributions}
I.A and H.L conceived and  conceptualised the research. I.A. did the analytical calculations and performed the numerical simulations. H.L. supervised the research. I.A. wrote the first draft, but both authors discussed and interpreted the results and wrote the final version of the manuscript.  

\subsection*{Competing interests}
The authors declare no competing interests. \\

\subsection*{Additional information}
\textbf{Supplementary information} The online version contains
supplementary material available at (the address will be provided by the journal) \\

\textbf{Correspondence} and requests for materials should be addressed to
Iman Abdoli.

\newpage
\cleardoublepage

\setcounter{page}{1}
\renewcommand{\thepage}{\arabic{page}}
\begin{titlepage}
	\centering
	{\large \textbf{Supplementary Material: Quadrupolar gyration of a Brownian particle in a confining ring} \par}
	\vspace{0.5cm}
{\large Iman Abdoli$^{1, \star}$ , Hartmut L\"{o}wen$^1$}\\[0.5cm]
{\small $^1$Institut für Theoretische Physik II - Weiche Materie, Heinrich-Heine-Universität Düsseldorf}\\[0.2cm]
{\footnotesize Email$^\star$: iman.abdoli@hhu.de}\\[0.2cm]
	\vfill
\end{titlepage}
\renewcommand{\theequation}{S\arabic{equation}}
\setcounter{equation}{0} 

\renewcommand{\thefigure}{S\arabic{figure}}
\setcounter{figure}{0} 
\vspace*{1.cm}	

	In this Supplementary Material, we provide the detailed derivations and supporting calculations of the theoretical results reported in the main text. It offers a transparent and self-contained mathematical framework of the methods used to investigate the  steady-state behaviour of a Brownian particle in a confining ring and driven out of equilibrium by anisotropic thermal noise–where
	the particle’s orthogonal spatial degrees of freedom are coupled to heat baths at unequal temperatures.
	
	We begin by deriving the time-dependent Fokker--Planck equation in polar coordinates from its Cartesian version. The resulting equation is then systematically analyzed using a narrow-ring expansion  to obtain perturbative expressions for the steady-state probability distribution, probability currents, and entropy production density. We then provide a dimensional analysis of all key observables. Finally, we summarize the Brownian dynamics simulation scheme used to validate our theoretical predictions and assess the range of validity of the perturbative approach. 

\section{Projection of the Fokker-Planck equation onto polar coordinates}
	
	\begin{figure}
		\vspace*{1.cm}	
		\centering
		\includegraphics[width=7.5cm]{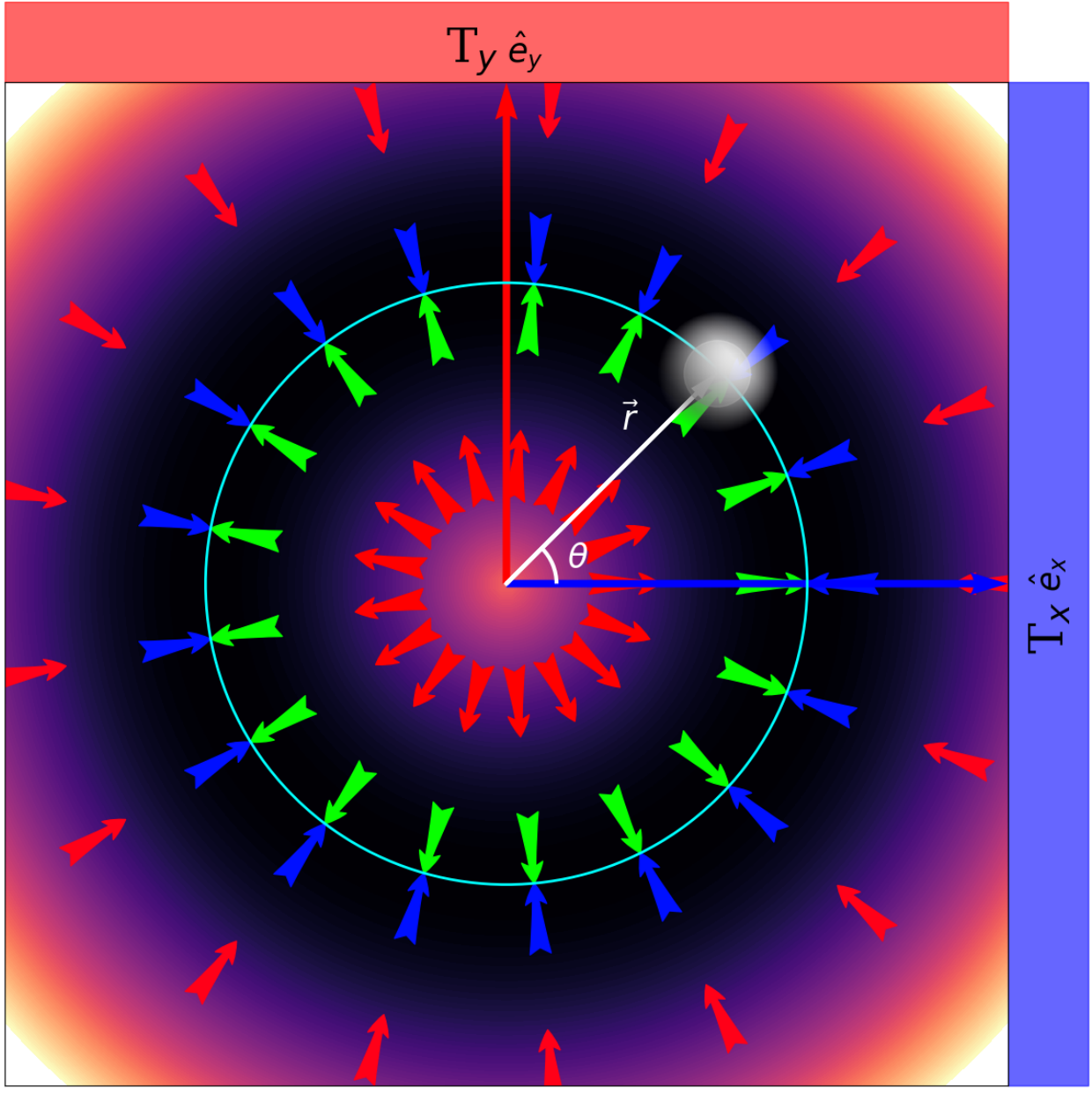}
		\caption{ \textbf{Schematic illustration of the model.} 
			Schematic of a ring Brownian gyrator. A Brownian particle (gray sphere) diffuses along a circular ring of radius $R$ (cyan) centered at the origin. The radial restoring force confines the particle to the ring, while the thermal environment imposes anisotropic noise along the Cartesian axes: the horizontal (blue) and vertical (red) bars represent the effective temperatures $T_x$ and $T_y$, respectively. The vector $\vec{r}$ indicates the instantaneous position of the particle, forming an angle $\theta$ with respect to the $x$-axis. The background color shows the radial potential landscape $U(r) =k (r - R)^2/2$, where $r$ is the distance from the origin and $k$ is the stiffness of the potential, with darker regions indicating lower potential energy and hence the preferred region of motion.} 
		\label{figure01_supp}
	\end{figure}
	
	We consider a Brownian particle confined in a ring  of radius $R$ modeled by the potential \(U(r) = \frac{k}{2} \left( r - R \right)^2\) where $r=\sqrt{x^2 + y^2}$. The particle is subjected to independent noise sources with distinct temperatures \(T_x\) and \(T_y\) along the orthogonal Cartesian axes $x$ and $y$ (see Fig.~\ref{figure01_supp}). The probability density of finding the particle at position \((x, y)\) at time \(t\), \(P(x, y, t)\) evolves according to the following Fokker-Planck equation~\cite{dotsenko2013two, chang2021autonomous, abdoli2022tunable}

	\begin{equation}
		\frac{\partial P(x, y, t)}{\partial t} = -\nabla \cdot \boldsymbol{J}(x, y, t),
	\end{equation}
	where the components of the probability flux \(\boldsymbol{J} = (J_x, J_y)\) are given by
	
	\begin{align}
		J_x(x, y, t) & = -\mu \frac{\partial U}{\partial x} P(x, y, t) - \mu T_x \frac{\partial P}{\partial x} \\
		J_y(x, y, t) &= -\mu \frac{\partial U}{\partial y} P(x, y, t) - \mu T_y \frac{\partial P}{\partial y}
	\end{align}
	where \(\mu\) is the mobility. We now aim to express the dynamics in polar coordinates $(r, \theta)$. Therefore, we project the Cartesian flux vector $\bm{J}(x, y)$ onto radial and angular directions, described by
	\begin{align}
		J_r &= \bm{J} \cdot \hat{\bm{e}}_r, \\
		J_\theta &= \bm{J} \cdot \hat{\bm{e}}_\theta,
	\end{align}
	where \(\hat{\bm{e}}_r = (\cos\theta, \sin\theta)\) and 
	\(\hat{\bm{e}}_\theta = (-\sin\theta, \cos\theta)\) are the local orthonormal basis vectors (or unit vectors).
	
	
	\vspace{1em}
	We start by the radial component of the probability flux \(J_r(r, \theta)\) where we project the Cartesian flux vector \( \bm{J}(x, y) = (J_x, J_y) \) onto the local radial direction  \( \hat{\bm{e}}_r \)
	\begin{align}
		J_r &= \hat{\bm{e}}_r \cdot \bm{J}(x, y) \nonumber \\
		&= -\mu \left( T_x \cos\theta \, \partial_x P + T_y \sin\theta \, \partial_y P \right) 
		- \mu k (r - R) P,
	\end{align}
	where the first two terms correspond to the diffusive fluxes in the \(x\) and \(y\) directions due to anisotropic noises, while the last term is the deterministic drift arising from the radial restoring force of the potential \(U(r)\). To re-express the above equation entirely in polar coordinates, we use the chain rule for the partial derivatives with respect to \(x\) and \(y\) to express in terms of \(r\) and \(\theta\)
	\begin{align}
		\label{S7}
		\partial_x &= \cos\theta \, \partial_r - \frac{\sin\theta}{r} \, \partial_\theta, \\
		\partial_y &= \sin\theta \, \partial_r + \frac{\cos\theta}{r} \, \partial_\theta,
		\label{S8}
	\end{align}
	
	These identities, given in Eqs.~\eqref{S7}--\eqref{S8}, can now be used to evaluate the diffusion terms explicitly in polar coordinates, which read

	\begin{align}
		T_x \cos\theta \, \partial_x P &= T_x \left( \cos^2\theta \, \partial_r P - \frac{\cos\theta \sin\theta}{r} \, \partial_\theta P \right), \\
		T_y \sin\theta \, \partial_y P &= T_y \left( \sin^2\theta \, \partial_r P + \frac{\sin\theta \cos\theta}{r} \, \partial_\theta P \right).
	\end{align}
	
	Note that the radial and angular derivatives are involved in both diffusion terms, reflecting the fact that anisotropic diffusion projects nontrivially onto both directions in polar coordinates.
	Combining terms, the full expression for the radial component of the flux can be written as
	
	\begin{align}
		\label{radial_flux_supp}
		J_r = - \left[ D_r(\theta) \, \partial_r P + \frac{ D_{r\theta}(\theta)}{r} \, \partial_\theta P + \mu k (r - R) P \right],
	\end{align}
	where we have defined the angle-dependent effective diffusion coefficients as
	
	\begin{align}
		D_r(\theta) &= \mu(T_x \cos^2\theta + T_y \sin^2\theta), \\
		D_{r\theta}(\theta) &= \frac{\mu}{2} (T_y - T_x) \sin(2\theta).
	\end{align}

	Following  similar projection steps, we can derive the angular component of the probability flux \(J_\theta(r, \theta)\) by projecting the Cartesian flux vector \(\bm{J}(x, y)\) onto the angular basis \(\hat{\bm{e}}_\theta \).  Applying the same coordinate transformations and grouping terms gives rise to a closed expression for \(J_\theta\) that explicitly reflects the angular dependence of the diffusion anisotropy. The final expression reads
	\begin{align}
		\label{angular_flux_supp}
		J_\theta  =  D_{r\theta}(\theta) \, \partial_r P 
		-  \frac{D_\theta(\theta)}{r} \, \partial_\theta P,
	\end{align}
	where we defined the angular diffusion coefficient
	\begin{align}
		D_\theta(\theta) &= \mu(T_x \sin^2\theta + T_y \cos^2\theta).
	\end{align}
	This completes the decomposition of the flux into polar components, given in Eq.~\eqref{radial_flux_supp} and Eq.~\eqref{angular_flux_supp}, and highlights how anisotropic thermal fluctuations induce off-diagonal coupling between radial and angular dynamics in the ring geometry.
	
	We can now write the time-dependent Fokker–Planck equation governing the evolution of the probability density \( P(r, \theta, t) \) in polar coordinates which reads
	
	\begin{equation}
		\label{eq:FPE_tensor}
		\frac{\partial P}{\partial t} = - \nabla \cdot \left[ \mu \mathbf{F} P - \mathbf{D}(\theta) \cdot \nabla P \right],
	\end{equation}
	where \(\mathbf{F} = -\nabla U(r) = -k(r - R)\, \hat{\mathbf{e}}_r\) is the radial restoring force derived from the ring-shaped potential. The gradient \(\nabla\) and divergence \(\nabla \cdot\) operators are understood in polar coordinates, where curvature contributions arise from the geometry. In this way, we derive the diffusion tensor \(\mathbf{D}(\theta)\)  in the polar basis as
	
	\begin{equation}
		\label{eq:Dtensor}
		\mathbf{D}(\theta) = \mu T 
		\begin{pmatrix}
			1 - \alpha \cos 2\theta & \alpha \sin 2\theta \\
			\alpha \sin 2\theta & 1 + \alpha \cos 2\theta
		\end{pmatrix},
	\end{equation}
	where \(\alpha = (\beta - 1)/(\beta+ 1)\) is the dimensionless anisotropy parameter and \(T = T_x(\beta + 1)/2\) is the mean effective temperature with \(\beta = T_y/T_x\) where \(\beta=1\) reduces the Fokker-Planck equation to the known equation for a single temperature system~\cite{Gardiner2009stochastics,balakrishnan2008elements}.  
	
	The expression, given in Eq.~\eqref{eq:Dtensor}, clearly show how the anisotropic thermal noises, although diagonal in Cartesian coordinates, induces angular modulations in the radial direction when projected into polar geometry giving rise to a \textit{tensorial diffusion coefficient}. The off-diagonal coupling \(D_{r\theta}(\theta)\) arises entirely due to the misalignment between the principal axes of the diffusion tensor and the local polar frame. 
	This formulation is exact and fully captures the time evolution of \( P(r, \theta, t) \) under anisotropic thermal fluctuations and confinement in a ring potential.
	
	\section{Steady-state solution: narrow ring}
	
	In this section, we analyze the steady-state behavior of the system by solving the Fokker--Planck equation derived in Eq.~\eqref{eq:FPE_tensor} and Eq.~\eqref{eq:Dtensor}. We develop a perturbative solution around the trap radius \( r = R \), treating the width of the ring as a formal expansion parameter. This allows us to systematically compute the steady-state probability density \( P(r, \theta) \), along with the corresponding radial and angular probability flux components \( J_r(r, \theta) \) and \( J_\theta(r, \theta) \). The resulting expressions reveal a quadrupolar symmetry in the flux field, characterized by oscillatory angular dependence in both current components. From these fluxes and the known anisotropic diffusion tensor, we derive the local entropy production density \( \sigma(r, \theta) \), which captures the spatial distribution of irreversible dissipation in the system. Angular averaging yields the radial profile \( \sigma(r) \), providing direct insight into the localization of nonequilibrium activity around the ring. All analytical results are obtained to leading order in the anisotropy parameter and are shown to quantitatively capture the emergence of quadrupolar gyration in the narrow-ring limit.
	
	\subsection*{Steady-state probability density function (PDF)}

	Our starting point is the Fokker--Planck equation given in Eq.~\eqref{eq:FPE_tensor}, and setting $\nabla\!\cdot\!\mathbf{J} = 0$, where the probability flux $\mathbf{J}$ depends explicitly on both the spatially varying diffusion tensor and the confining potential. Assuming the particle is strongly localized near the ring radius $R$ (i.e. a narrow ring), the stationary distribution can be approximated by a Gaussian radial profile centered at $r=R$, with an angular modulation induced by the temperature difference along orthogonal Cartesian directions.
	
	Therefore, we consider the following ansatz for the stationary probability density
	
	\begin{align}
		\label{P0_ansatz}
		P_0(r,\theta) &= 
		\frac{1}{\sqrt{T_r(\theta)}} 
		\exp\!\left[-\frac{k (r-R)^2}{2 T_r(\theta)}\right],\\
		P(r,\theta) &= P_0(r,\theta)\,
		\exp\!\bigl[-\alpha\, g(r)\cos(2\theta)\bigr],
		\label{P_ansatz}
	\end{align}
	where $P_0$ captures the leading-order Gaussian localization in the radial direction where we define the radial effective temperature as $T_r(\theta)  = T (1-\alpha\cos(2\theta))$. The second exponential factor in Eq.~\eqref{P_ansatz} represents an angular modulation that accounts for symmetry breaking due to the temperature difference, which is encoded in the anisotropy parameter $\alpha$.
	
	We now aim to determine the function $g(r)$, for which we plug the ansatz into the steady-state Fokker--Planck equation and expand to linear order in $\alpha$, consistently neglecting higher-order terms $\mathcal{O}(\alpha^2)$ and beyond. Matching the coefficient of $\cos(2\theta)$ at each order results in a linear differential equation for $g(r)$ whose solution provides us with the angular correction to the stationary distribution. The differential equation for $g(r)$ reads

	\begin{equation}
		r^2 T g''(r)\!+\!r\bigl[ T\! -\! k r(r\! -\! R)\bigr]g'(r)\!  =\! -\!2\left(k(r\!-\!R)R\!+\!T\right).
	\end{equation}
	
	
	We now assume that $r$ is close to $R$, i.e., we set
	\(r = R + \rho, \quad |\rho| \ll R\)
	and expand
	\( r^2 \approx R^2 + 2R\rho\).
	Thus, our ordinary differential equation (ODE) becomes
	\begin{align}
		& \bigl[R  T + \rho ( T - k R^2)\bigr] g'[r] 
		+  (R^2 + 2R\rho) T \, g''[r] \notag \\
		& = -2kR\rho - 2 T.
	\end{align}
	
	We approximate $g(R+\rho)$ as a polynomial expansion
	\[
	g(R+\rho) \approx a_0 + a_1 \rho + a_2 \rho^2,
	\]
	and thus,
	\[
	g'(R+\rho) \approx a_1 + 2a_2\rho, \qquad g''(R+\rho) \approx 2a_2.
	\]
	Expanding and matching the constant and $\rho$ terms, will give us the following set of linear equations   
	
	\begin{equation}
		\begin{cases}
			R \, a_1 + 2R^2 \, a_2 = -2, \\ 
			( T - k R^2) \, a_1 + 4R  T \, a_2 = -2kR.
		\end{cases}
	\end{equation}
	
	These linear equations determine $a_1$ and $a_2$. Solving, we obtain
	\begin{align}
		\label{a1_supp}
		a_1 &= \frac{2k R^2 - 4 T }{k R^3 + R T },  \\
		a_2 &= \frac{-2k R^2 + T }{k R^4 + R^2 T }. 
		\label{a2_supp}
	\end{align}
	
	Note that considering the characteristic length scale as \(\sqrt{T_x/k}\) ,  \(a_1\) and \(a_2\) have the units of \(1/L\) and \(1/L^2\), respectively.

	The final approximate steady-state probability density reads
	\begin{equation}
		\label{pdf_per}
		P(r,\theta) \approx \frac{1}{\mathcal{N}\sqrt{T_r(\theta)}} e^{-\frac{k\rho^2}{2 T_r(\theta)}}e^{-\alpha\,(a_1 \rho + a_2 \rho^2)\cos(2\theta)}.
	\end{equation}
	Here, $a_1$ and $a_2$ are given by the explicit expressions above, and $\rho = r - R$. To determine the full steady-state probability density, we also require the normalization factor $\mathcal{N}$. 
	In the narrow-ring approximation, the radial profile is sharply peaked around $r = R$ and can be approximated by a Gaussian of width $\sqrt{T_r(\theta)/k}$. 
	Integrating over $r$ then yields a factor $\sqrt{2\pi T_r(\theta)/k}$, while the angular integral contributes $2\pi$ to leading order. 
	Therefore, to lowest order in $a$ and for a narrow ring, the total normalization factor is approximately
	\[
	\mathcal{N} \approx 2\pi R \sqrt{\frac{2\pi}{k}},
	\]
	which ensures that $P(r, \theta)$ integrates to one. 
	This confirms that, to first approximation, the radial confinement dominates the normalization, and angular corrections enter only perturbatively. \\

	\subsection*{Leading-order fluxes in the narrow-ring approximation}
	
	Starting from the approximate steady-state probability density in Eq.\eqref{pdf_per}, we proceed to evaluate the radial and angular probability currents.  The radial and angular flux components are given by
	\begin{align}
		J_r &= -D_r(\theta)\frac{\partial P}{\partial r} - D_{r\theta}(\theta)\frac{1}{r}\frac{\partial P}{\partial \theta} - \mu k(r - R)P, \\
		J_\theta &= -D_\theta(\theta)\frac{1}{r}\frac{\partial P}{\partial \theta} - D_{r\theta}(\theta)\frac{\partial P}{\partial r}.
	\end{align}
	For a narrow ring (large $k$), the radial profile is strongly localized around $r = R$, allowing us to keep only the leading-order contributions in $a$ and $\rho$.
	
	Using the approximate forms and keeping only terms linear in $\alpha$ (neglecting $\alpha^2$ and higher), we obtain explicit simplified expressions for $J_r$ and $J_\theta$. To leading order, these are given by
	\begin{align}
		\label{radial_flux}
		J_r(r,\theta) 
		&\approx \frac{\alpha\mu \sqrt{kT}\,e^{-\frac{k(r-R)^2}{2T}}}{(2 \pi)^{3/2}R}
		\left(a_1 + 2 a_2 (r - R)\right) \,\cos(2\theta), 
		\\
		J_\theta(r,\theta) 
		&\approx \frac{\alpha\mu\sqrt{k}\,e^{-\frac{k(r-R)^2}{2\mu T}}}{\sqrt{T}(2 \pi)^{3/2}R\,r}
		\Bigl(T + \notag \\
		\quad & (r - R)\bigl[k R - 2 (a_1 + a_2 (r - R))T \bigr]\Bigr)
		\sin(2\theta).
		\label{angular_flux}
	\end{align}
	Here, $a_1$ and $a_2$ are explicitly known from the narrow-ring expansion, which are given in Eq.\eqref{a1_supp} and Eq.\eqref{a2_supp}, respectively.  These closed-form flux expressions explicitly show the angular symmetry-breaking contributions driven by the small anisotropy parameter $\alpha$, while maintaining explicit dependence on the ring parameters $k$, $R$, and $T$. 
	The exponential prefactor $\exp\left[-k(r-R)^2/2\mu T\right]$ ensures confinement around $r=R$, confirming the strong localization of both the probability density and the associated fluxes. This completes the leading-order perturbative analysis of the flux structure in the narrow-ring approximation.
	
	\begin{figure*}[t]
		\centering
		\includegraphics[width=16.5cm]{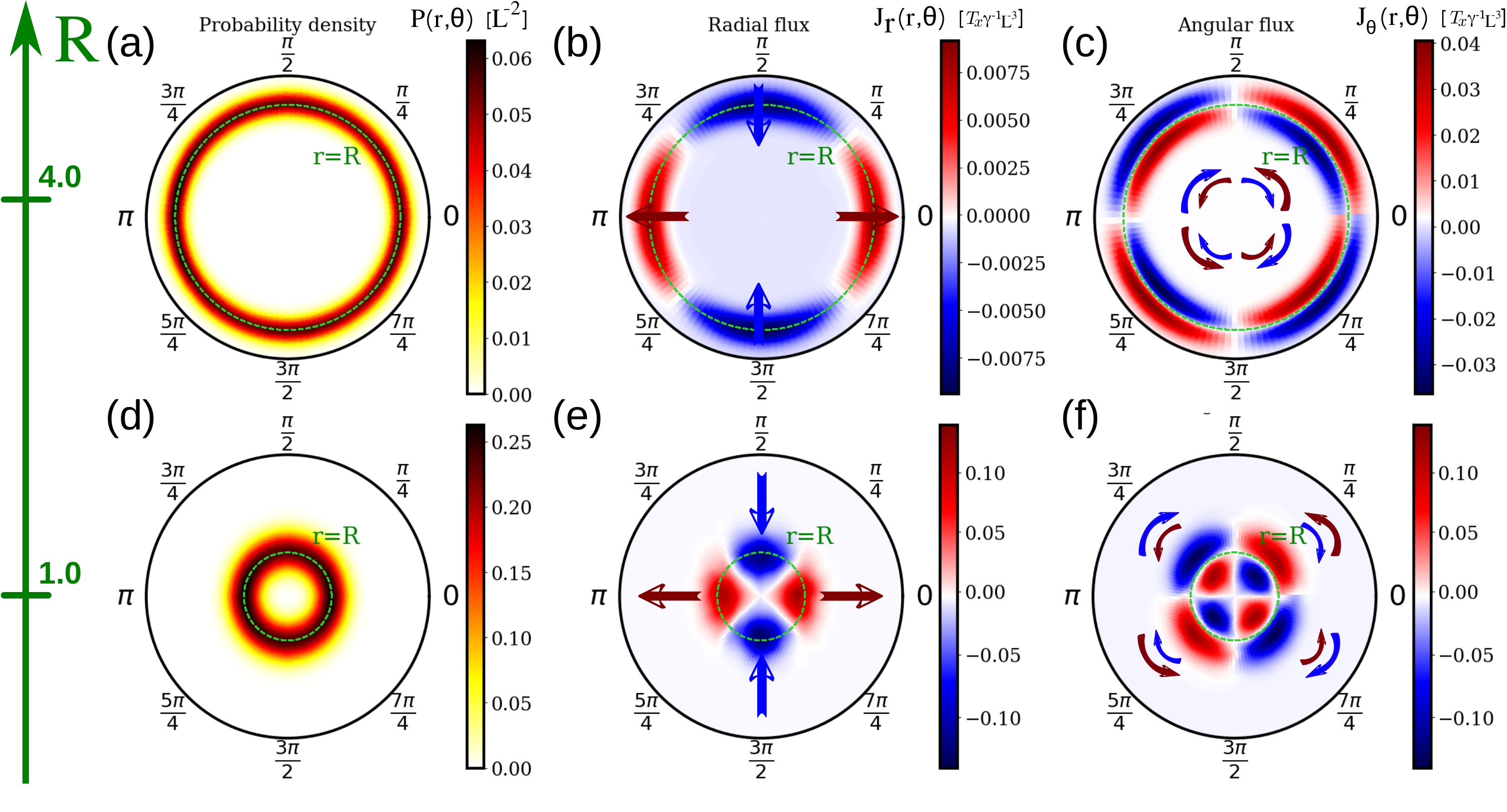}
		\caption{
			\textbf{Simulations: Steady-state probability density and probability flux components.}
			Panels (a–c) correspond to \(R = 4.0\), while (d–f) show \(R = 1.0\). In all cases, the trap stiffness is \(k = 20\), and the temperature ratio \(\beta=1.5\), with characteristic length scale \(L = \sqrt{T_x / k}\).	
			(a,d) The probability distribution is sharply localized around the ring \(r = R\) (indicated by dashed circles), and anisotropic thermal noise induces an angular modulation visible as slight distortions along the \(\theta\)-direction.	
			(b,e) The radial flux highlights the local circulating currents that maintain the non-equilibrium steady state. Inward and outward flow regions are indicated by blue and dark-red straight arrows, respectively, illustrating the direction of radial motion.	
			(c,f) The angular flux emphasizes the sustained azimuthal circulating motion characteristic of gyrators. Blue and dark-red curved arrows indicate clockwise  and counter-clockwise  circulations, respectively.	
			For \(R = 4.0\) (top row), the probability density forms a broad annulus with smoother flux patterns, indicating a more delocalized state. In contrast, for \(R = 1.0\) (bottom row), the probability is tightly localized and the flux components are more concentrated, highlighting stronger confinement and enhanced rotational features.	
			The dashed circles at \(r = R\) serve as a reference to the nominal ring radius and emphasize the degree of spatial localization. The results are computed from numerical simulations of Eq.~\eqref{langevin}, which are in an excellent agreement with Fig.$2$ of the main text. 
		}
		
		\label{figure02_supp}
	\end{figure*}
	\subsection*{Entropy production density}
	
	In overdamped systems with anisotropic thermal noise, local irreversibility can be quantified by the steady-state entropy production density, defined as
	\begin{equation}
		\sigma(\mathbf{x}) 
		= \frac{\mathbf{J}^\top(\mathbf{x}) \, \mathbf{D}^{-1}(\mathbf{x}) \, \mathbf{J}(\mathbf{x})}{P(\mathbf{x})},
	\end{equation}
	where $\mathbf{J}(\mathbf{x})$ is the steady-state probability flux, $\mathbf{D}(\mathbf{x})$ is the diffusion tensor, and $P(\mathbf{x})$ is the steady-state probability distribution.
	
	In our model, we work in polar coordinates $\mathbf{x} = (r,\theta)$, with the flux vector
	\begin{equation}
		\mathbf{J}(r,\theta) 
		= 
		\begin{pmatrix}
			J_r(r,\theta) \\
			J_\theta(r,\theta)
		\end{pmatrix},
	\end{equation}
	and the angular-dependent diffusion tensor expressed as
	\begin{equation}
		\mathbf{D}(\theta)
		=
		\mu T
		\begin{pmatrix}
			1 - \alpha \cos 2\theta & \alpha \sin 2\theta \\
			\alpha \sin 2\theta & 1 + \alpha \cos 2\theta
		\end{pmatrix},
	\end{equation}
	where $\alpha = (\beta - 1)/(\beta + 1)$ is the dimensionless noise anisotropy and $T = T_x (\beta + 1)/2$ the average temperature. The determinant of this tensor simplifies to
	\begin{equation}
		\det \mathbf{D} = (\mu T)^2 (1 - \alpha^2),
	\end{equation}
	which is constant and independent of $\theta$, enabling closed-form analytical solutions.
	
	The inverse diffusion tensor takes the standard form
	\begin{equation}
		\mathbf{D}^{-1}(\theta)
		= \frac{1}{\det \mathbf{D}} \,
		\begin{pmatrix}
			D_\theta(\theta) & -D_{r\theta}(\theta) \\
			-D_{r\theta}(\theta) & D_r(\theta)
		\end{pmatrix}.
	\end{equation}
	Using this structure, the entropy production density becomes
	\begin{equation}
		\sigma(r,\theta)
		= \frac{J_r^2 D_\theta - 2 J_r J_\theta D_{r\theta} + J_\theta^2 D_r}
		{(\mu T)^2 (1 - \alpha^2) P(r,\theta)}.
	\end{equation}
	
	At leading order in the narrow-ring limit, the steady-state probability and flux fields admit approximate analytical expressions, yielding a closed-form for the local entropy production density:
	\begin{align}
		&	\sigma(r,\theta) 
		= \notag \\ 
		& \frac{\alpha^2 \, e^{- \frac{k (r - R)^2}{2T}} \, \sqrt{k} \, \mu }
		{4 \sqrt{2}\,\pi^3 \, T^{3/2} \, (R - R \alpha^2)}
		\Bigl[
		\bigl(a_1 + 2 a_2 (r - R)\bigr)^2 T^2 \cos^2(2\theta)
		\nonumber\\
		&
		+ \frac{\bigl(T + (r - R)\bigl(kR - 2(a_1 + a_2 (r - R))T\bigr)\bigr)^2 \sin^2(2\theta)}
		{r^2}
		\Bigr].
	\end{align}
	which exhibits a pronounced quadrupolar angular structure encoded by $\cos^2(2\theta)$ and $\sin^2(2\theta)$ modulations, and a radial Gaussian envelope centered around $r = R$.
	
	The corresponding radial profile, obtained by angular averaging, reads
	\begin{align}
		\sigma(r) =\ 
		& \frac{\alpha^2 e^{-\frac{k (r - R)^2}{2 T}} \sqrt{k} \mu}
		{8 \sqrt{2} \pi^4 T^{3/2} (R - R \alpha^2)} \left[
		\pi (a_1 + 2 a_2 (r - R))^2 T^2 \right. \notag \\
		& \left. +\ \frac{\pi \left( T + (r - R)(k R - 2(a_1 + a_2 (r - R)) T) \right)^2}{r^2}
		\right],
	\end{align}
	revealing nontrivial radial dependence with peaks that shift and grow with increasing anisotropy $\alpha$. The amplitude of $\sigma(r)$ scales as $\alpha^2$, highlighting that entropy production is entirely driven by the imposed temperature gradient. This structure reflects the interplay between geometric confinement, flux topology, and anisotropic fluctuations.

	\section*{Dimensional analysis of key quantities}
	
	In this section, we discuss the dimensionality of fundamental quantities that characterize nonequilibrium steady states in our system: the probability density function, the probability flux and the entropy production density. Understanding their units is essential for interpreting numerical values, comparing different parameter regimes, and properly scaling simulation or experimental results.

	We express the stationary probability density in a dimensionless form by rescaling lengths and temperatures using the characteristic length scale \( L = \sqrt{T_x/k} \) and the reference temperature \( T_x \). We define the dimensionless variables
	\(
	\tilde{r} = r/L, \quad 
	\tilde{R} = R/L, \quad
	\tilde{\rho} = \tilde{r} - \tilde{R}, \quad
	\tilde{T}_r(\theta) = \frac{T_r(\theta)}{T_x} 
	= \tilde{T}\bigl(1-\alpha\cos(2\theta)\bigr)\) where \(\tilde{T} = T/T_x = (1+\beta)/2\).
	Substituting these definitions into the approximate stationary solution obtained above yields the expression
	\begin{align}
		P(r,\theta) 
		&\approx 
		\frac{1}{L^2 \tilde{R} \sqrt{\tilde{T}_r(\theta)} (2\pi)^{3/2}} \times \notag\\
		&\exp\!\Bigl[-\frac{\tilde{\rho}^2}{2 \tilde{T}_r(\theta)}
		- \alpha\,(\tilde{a}_1 \tilde{\rho} + \tilde{a}_2 \tilde{\rho}^2)\cos(2\theta)\Bigr],
	\end{align}
	where the coefficients \(\tilde{a}_1 = a_1 L\) and \(\tilde{a}_2 = a_2 L^2\) are the dimensionless counterparts of the parameters \(a_1\) and \(a_2\) given in Eq.\eqref{a1_supp} and Eq.\eqref{a2_supp}. Therefore, the dimensionless probability density can be written as $\tilde{P}(\tilde{r}, \theta) = L^2 P(r, \theta)$.

	Using the same dimensionless variables, we can derive the expressions for the radial and angular fluxes providing proper units of the fluxes. Plugging the dimensionless variables into Eq.~\eqref{radial_flux} and Eq.~\eqref{angular_flux} gives
	
	\begin{align}
		J_r(r,\theta) 
		&\approx \frac{\mu T_x\alpha \sqrt{\tilde{T}}\,e^{-\frac{\tilde{\rho}^2}{2\tilde{T}}}}{(2 \pi)^{3/2}L^3}
		\left(\tilde{a}_1 + 2 \tilde{a}_2 \tilde{\rho}\right) \,\cos(2\theta), 
		\\
		J_\theta(r,\theta) 
		&\!\approx\! \frac{\mu T_x\alpha\,e^{-\frac{\tilde{\rho}^2}{2\tilde{ T}}}}{\sqrt{\tilde{T}}(2 \pi)^{3/2}\tilde{R}\,\tilde{r}L^3}
		\!\Bigl[\tilde{T} \!\!+\!\!  \tilde{\rho}\bigl[\tilde{R}\!\! -\!\! 2 (\tilde{a}_1\! \!+\!\! \tilde{a}_2 \tilde{\rho})\tilde{T} \bigr]\!\Bigr]
		\!\sin(2\theta),
	\end{align}
	where the dimensionless expressions for the radial and angular fluxes can be written as 
	
	\begin{align}
		\tilde{J}_r(\tilde{r},\theta) & = \frac{L^3}{\mu T_x} J_r(r,\theta) \\
		\tilde{J}_\theta(\tilde{r},\theta) & = \frac{L^3}{\mu T_x} J_\theta(r,\theta).
	\end{align}
	
	A similar calculation will give the unit of the entropy production density as \(L^4/\mu T_x\). We summarize these results compactly, we realize the following units 
	
	\begin{align}
		[P] & = L^{-2}, \\
		[J] &= L^{-1}\,t^{-1},\\
		[\sigma] &= L^{-2}\,t^{-1},
	\end{align}
	
	where time $t$ has the unit of $L^2/\mu T_x$. 

	\section{Brownian Dynamics Simulations}
	\begin{figure}[t]
		\centering
		\includegraphics[width=8.0cm]{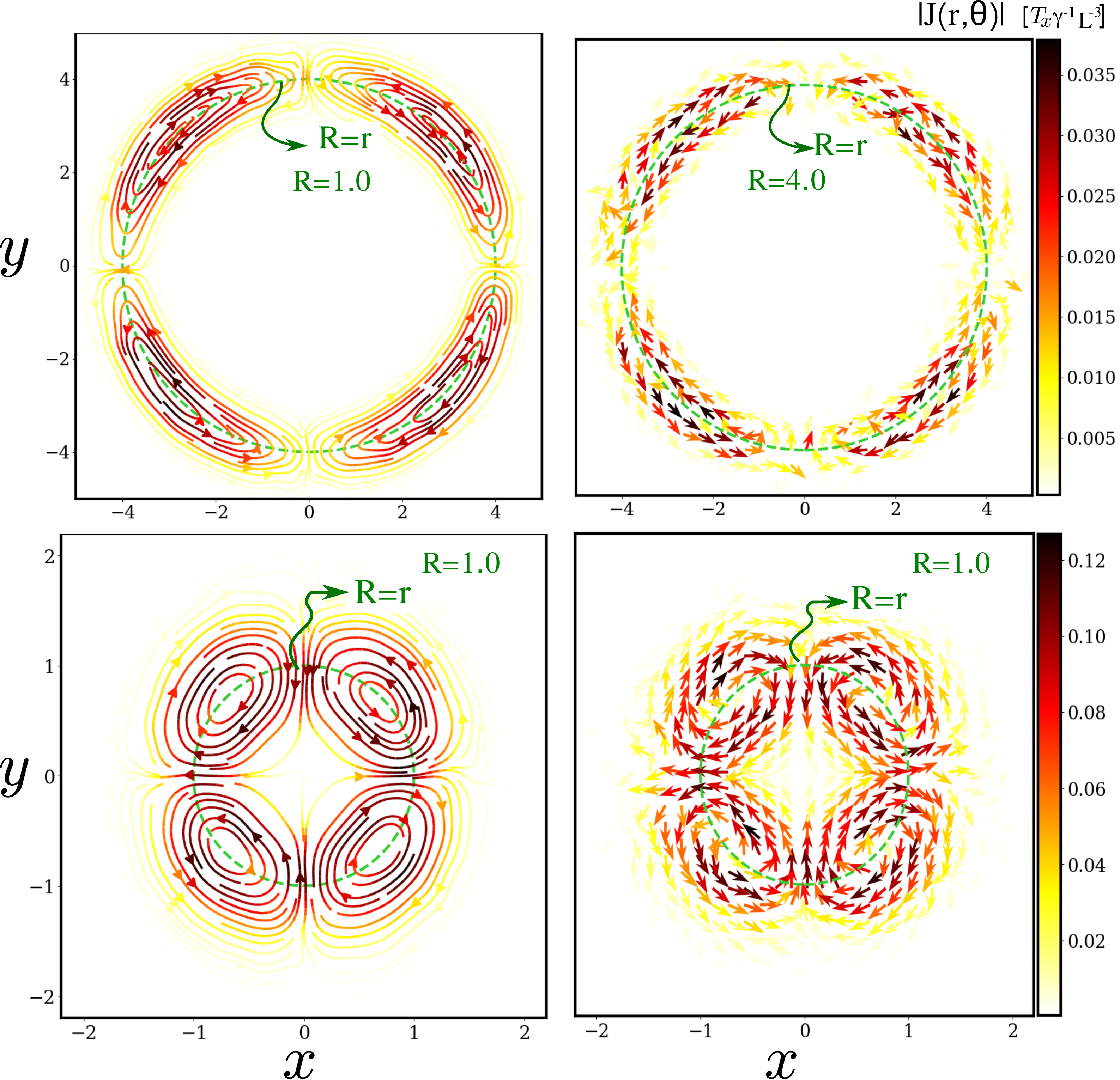}
		\caption{
			\textbf{Simulations: Probability flux streamlines and quiver plots.}
			Top panels correspond to $R = 4.0$, bottom panels to $R = 1.0$. The steady-state probability flux $\mathbf{J}(r,\theta)$ is projected in Cartesian coordinates $(x, y)$, with streamlines visualizing circulation patterns and quiver arrows indicating local flux directions. Color maps encode the flux magnitude $|\mathbf{J}(r,\theta)|$ for \(\beta=1.5\) and \(k=20\), with $L = \sqrt{T_x/k}$. For $R = 4.0$ (top), the flux forms extended vortex-like loops, while for $R = 1.0$ (bottom), circulation becomes more localized with pronounced radial and azimuthal components due to tighter confinement and stronger anisotropy. Green dashed circles mark the nominal ring position $r = R$.  The results are computed from numerical simulations of Eq.~\eqref{langevin}, which are in an excellent agreement with Fig.$3$ of the main text.
		}
		
		\label{figure03_supp}
	\end{figure}

	In this section, we provide the details of numerical simulations for the verification of our theoretical predictions presented in the main text. We next present results for parameter values beyond the validity of our theory.  The dynamics of the particle's position \(\bm{r}(t) = (x(t), y(t))\) can be described by the following overdamped Langevin equation
	\begin{figure*}[t]
		\centering
		\includegraphics[width=16.5cm]{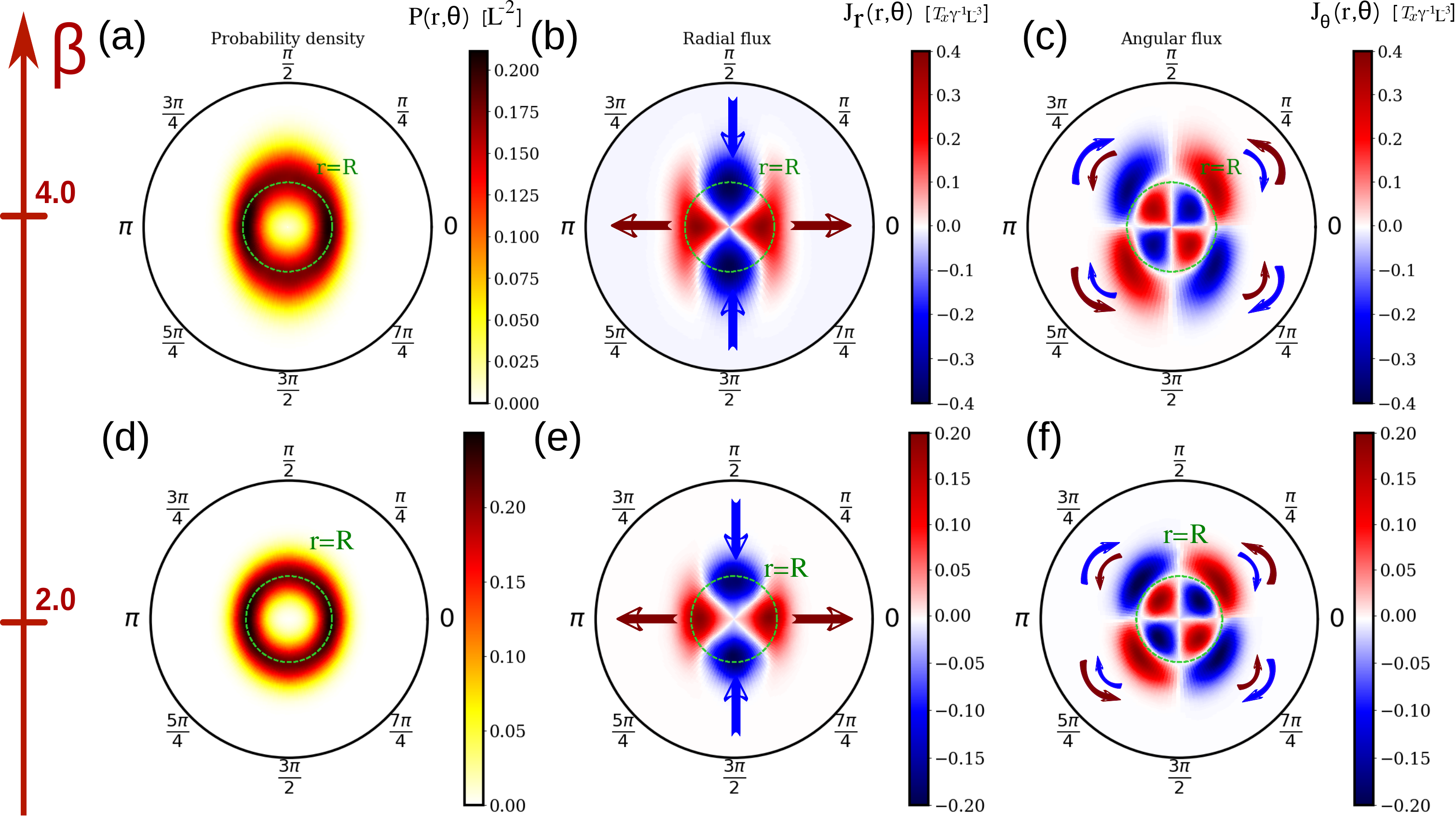}
		\caption{
			\textbf{Simulations: Steady-state probability density and probability flux components.}
			Panels (a–c) correspond to the temperature ratio \(\beta = 4.0\), while (d–f) show \(\beta = 2.0\). In all cases, the trap stiffness is \(k = 20\), and radius of the ring \(R = 1.0\), with characteristic length scale \(L = \sqrt{T_x / k}\).	
			(a,d) The probability distribution is sharply localized around the ring \(r = R\) (indicated by dashed circles) more stretched along hot axis.	
			(b,e) The radial flux, $J_r(r, \theta)$ highlights the local circulating currents that maintain the non-equilibrium steady state. Inward and outward flow regions are indicated by blue and dark-red straight arrows, respectively, illustrating the direction of radial motion.	
			(c,f) The angular flux, $J_\theta(r, \theta)$ emphasizes the sustained azimuthal circulating motion characteristic of gyrators. Blue and dark-red curved arrows indicate clockwise  and counter-clockwise  circulations, respectively.		
			The dashed circles at \(r = R\) serve as a reference to the nominal ring radius and emphasize the degree of spatial localization. The results are computed from numerical simulations of Eq.~\eqref{langevin}.
		}
		
		\label{figure04_supp}
	\end{figure*}
	
	\begin{align}
		\label{langevin}
		\dot{\bm{r}} = -\mu \bm{\nabla} U(r) + \bm{\xi}(t)
	\end{align}
	where \(\mu\) is the mobility, \(\boldsymbol{\xi}(t) = (\xi_x(t), \xi_y(t))\) is a Gaussian white noise vector with zero mean and \(\langle \eta_i(t)\eta_j(t') \rangle = 2\mu T_i \delta_{ij} \delta(t - t')\) where \(i,j \in \{x, y\}\)

	The simulations are performed using an Euler-Maruyama integration scheme with a discrete time step \(d t=1\times 10^{-3}\gamma/k\), chosen to be sufficiently small to ensure numerical stability and accuracy. Thermal noise is implemented by drawing random numbers from a normal distribution at each time step.  \\

\subsection*{Verification of the theoretical predictions}
We first show that for all parameter sets considered in the main text, the results of numerical simulations are in excellent agreement with the theoretical predictions derived in the narrow-ring approximation. The comparison, shown in Figs.~\ref{figure02_supp}--\ref{figure03_supp}, confirms that the perturbative solution accurately captures both the spatial structure and the magnitude of all steady-state quantities. Within the regime of validity of the approximation, the theory reproduces the simulation results, thereby validating our analytical results.

	\begin{figure}[b]
		\centering
		\includegraphics[width=8.0cm]{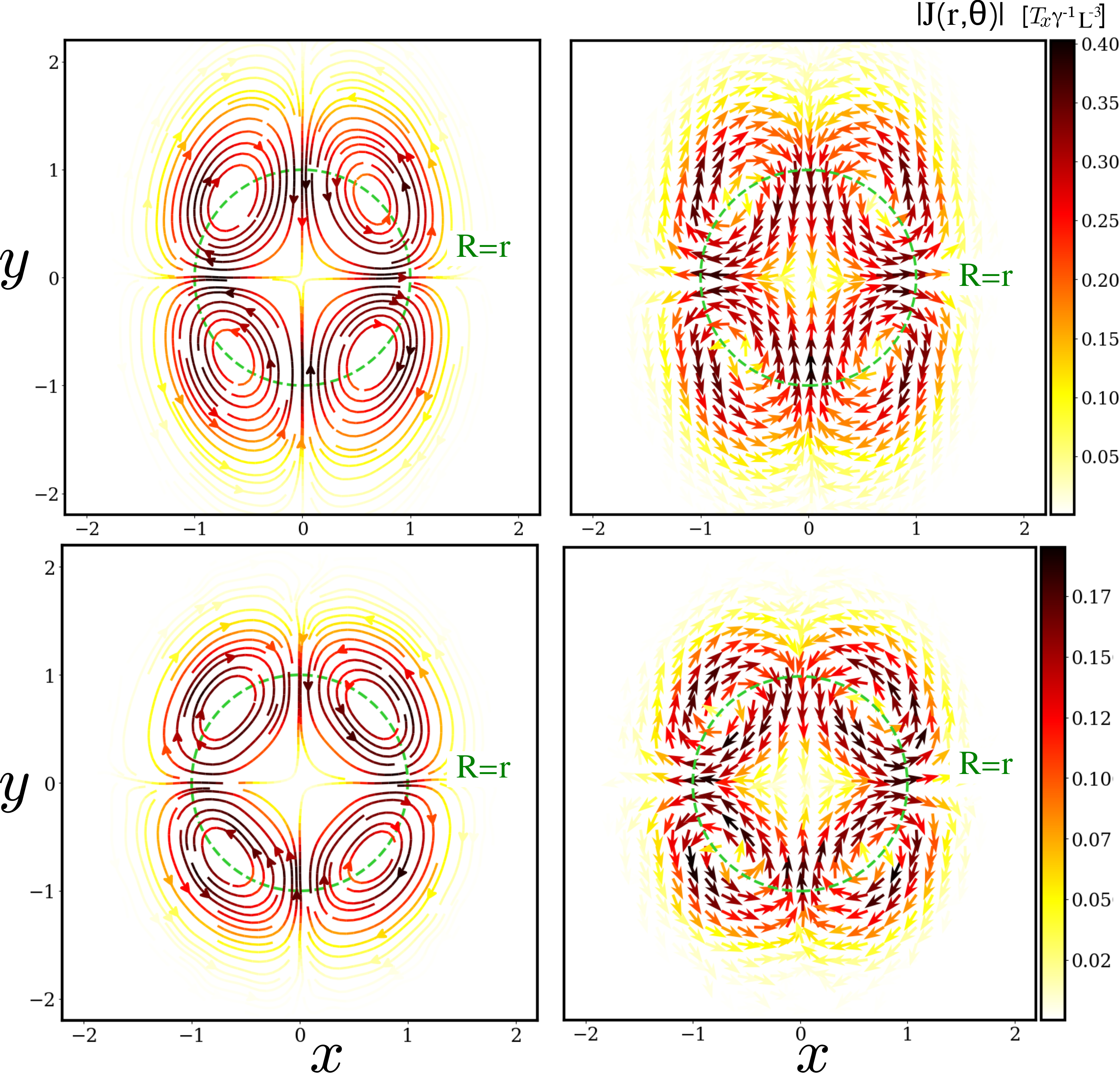}
		\caption{
			\textbf{Simulations: Probability flux streamlines and quiver plots}
			Top panels correspond to $\beta = 4.0$, bottom panels to $\beta = 2.0$ for the same system as presented in Fig.~\ref{figure04_supp}. The steady-state probability flux $\mathbf{J}(r,\theta)$ is projected in Cartesian coordinates $(x, y)$, with streamlines visualizing circulation patterns and quiver arrows indicating local flux directions. Color maps encode the flux magnitude $|\mathbf{J}(r,\theta)|$ in units of $[T_x/(\gamma L^3)]$, with $L = \sqrt{T_x/k}$. The results are computed from numerical simulations of Eq.~\eqref{langevin}.
		}
		
		\label{figure05_supp}
	\end{figure}
	
	\subsection*{Beyond theoretical predictions}
	To assess the limits of validity of our theoretical results, we perform Brownian dynamics simulations for larger temperature ratios $\beta = T_y/T_x$, where the perturbative expansion in the anisotropy parameter $\alpha$ is no longer strictly applicable. While the overall qualitative features, such as the quadrupolar flux structure, remain visible, quantitative deviations become apparent at $\beta \gtrsim 2$. Figures~\ref{figure04_supp} and~\ref{figure05_supp} represent the the probability density and probability fluxes from the numerical simulations for $\beta=2$ and $\beta=4$. The results clearly show that with increasing temperature ratio the anisortopy is more pronounced as can be seen in the radially stretched probability densities along the hot axis in Fig.~\ref{figure04_supp} (a, d). This is followed by a stretched radial flux along vertical direction where the temperature is higher, as showon in Fig.~\ref{figure04_supp} (b, e). In combination with the angular fluxes (Fig.~\ref{figure04_supp} (c, f)) the overall quadrupolar fluxes are pronounced as evident in corresponding steam and vector plots in Fig.~\ref{figure05_supp}.

	\section*{References}
	
	%

\end{document}